\documentclass[preprint,
preprintnumbers,amsmath,amssymb]{revtex4}
\usepackage{graphicx}% Include figure files
\usepackage{dcolumn}% Align table columns on decimal point
\usepackage{bm}% bold math

\def\beq{\begin{equation}}
\newcommand{\be}{\begin{equation}}
\newcommand{\ee}{\end{equation}}
\newcommand{\ba}{\begin{eqnarray}}
\newcommand{\ea}{\end{eqnarray}}
\begin{document}

%\preprint{APS/123-QED}

\author{Sandip Biswas and Kirill Melnikov
        \thanks{e-mail: kirill@phys.hawaii.edu}}
\affiliation{Department of Physics and Astronomy,
          University of Hawaii,\\ 2505 Correa Rd. Honolulu, HI 96822}  

\begin{flushright}
\vbox{
\begin{tabular}{l}
UH-511-1098-06\\
 hep-ph/0611345
\end{tabular}
}
\end{flushright}

\title{The rotation of 
magnetic field does not impact vacuum birefringence }

\begin{abstract}
We study how  the rotation of a classical magnetic 
field influences vacuum birefringence. In spite of the 
fact that solutions of the wave equation depend 
in a non-trivial way on the relative magnitude of the strength of vacuum 
birefringence and the angular rotation frequency of the 
magnetic field,  the induced ellipticity of the light ray 
does not depend on the magnetic field angular rotation frequency.
Hence, in contrast to recent claims, the result of the PVLAS experiment
cannot be explained by conventional non-linear quantum electrodynamics.
\end{abstract}

                              %display desired
\maketitle

\section{Introduction}
Because of quantum effects, photons can interact with each other.
For photon energies  smaller than the electron mass,
these quantum effects are encapsulated in the Euler-Heisenberg 
Lagrangian \cite{eh} and lead to a host of interesting phenomena such as 
photon splitting and vacuum birefringence in 
the presence of a classical electromagnetic  field \cite{adler1}.

The vacuum birefringence and vacuum dichroism 
are important effects  to search for 
hypothetical light particles such as axions -- neutral spin zero 
bosons, that  mix with photons in the presence of a classical 
magnetic field.
Recently,  the PVLAS experiment \cite{pvlas}
claimed to observe these effects; the PVLAS signal  
exceeds QED expectations by a factor $10^{4}$.
While the PVLAS claim is quite interesting, it is not easy to 
reconcile parameters of an axion that would lead to such an enhancement
with existing  constraints that 
mostly come from astrophysical observations. A number of models  that
 explain how this can be accomplished 
were recently discussed 
 in the literature
\cite{models}.

On the other hand, the setup of the PVLAS experiment is somewhat peculiar.
While 
description  of 
a photon propagation
 in  the external
magnetic field was developed  long ago \cite{adler1}, 
it applies to  the case of {\it static} magnetic field. Strictly speaking, 
these results can not be used to describe  the PVLAS 
experiment where the external magnetic field {\it rotates}.
This difference motivated the re-analysis of the photon propagation 
in the external rotating magnetic field and resulted in a recent 
claim \cite{port} that the effects observed in the PVLAS experiment 
can be described by conventional non-linear quantum electrodynamics. Intrigued 
by this claim, we decided to study the impact of  the rotation 
of the magnetic field  on the photon propagation. 
Our findings do not support the claim 
in Ref.\cite{port} and rule out the explanation of the PVLAS results 
based on non-linear quantum electrodynamics.

\section{Equations of motion}

As  stated earlier, the dynamics of photons with energies smaller than 
the mass of the electron is described by the Lagrangian \cite{eh,ll}
\be
{\mathcal{L}} = {\frac{1}{8{\pi}}}\left(\mathbf{E}^2 -\mathbf{B}^2\right) +
\frac{\xi}{8\pi}
\left[ {\left(\mathbf{E}^2 -\mathbf{B}^2\right)}^2 + 7{\left(\mathbf{E B}\right)}^2\right],
\label{eq1}
\ee
where $\mathbf{E},\mathbf{B}$ are the electric and magnetic fields,
$\displaystyle \xi = \frac{\alpha^2}{45~\pi m_e^4}$
 and  $\alpha$ and $m_e$ are the fine structure constant and the 
mass of the electron, respectively.

We would like to apply the Lagrangian Eq.(\ref{eq1}) to describe 
propagation of a photon with the wave vector ${\bf k}$
in the external rotating magnetic field $B_0$. Suppose the magnetic field 
rotates with the angular frequency $\omega_0$. 
For the PVLAS setup \cite{pvlas},
$\omega_0/k \ll 1$ and $ \xi B_0^2 \ll 1$. 
We work to first order in both of these parameters.  

To describe the photon propagation in the external magnetic field, we need 
to linearize the Lagrangian Eq.(\ref{eq1}) with respect to photon 
electric and magnetic fields. To this end, 
we write the electric and magnetic fields as superpositions of 
classical and quantum fields,
\begin{displaymath}
\mathbf{E} = {\mathbf{E}}^q,\;\;\;\;\;\;{\bf B} 
= {\mathbf{B}}^c +{\mathbf{B}}^q.
\end{displaymath}
We point out that the rotating magnetic field 
induces the electric field; so  both classical 
fields ${\mathbf{E}}^c$ and 
${\mathbf{B}}^c$ are required, in principle.
However, since the rotation frequency of 
the magnetic field is small, the influence of the induced electric field 
on the photon propagation is  a higher order effect with respect to 
$\omega_0/k$ and $\xi B_0^2$ and we neglect it in what follows.

Linearizing Eq.(\ref{eq1}) with respect to ${\mathbf{B}}^q$ and introducing 
the vector potential 
$A^{q,\mu} = (0, \mathbf{A}^q)$, we derive the equation of motion 
for the photon field
\be
\left({{\partial}_0}^2 - {{\vec \partial}}^2\right){\mathbf{A}}^q = \hat \Lambda\mathbf{A}^q.
\label{eq2}
\ee
The matrix $\hat \Lambda$ is specified below. We point out that 
in deriving Eq.(\ref{eq2}) we neglected all time derivatives of the 
classical magnetic field since they introduce effects that are suppressed 
by additional powers of $\omega_0/k$ and $\xi B_0^2$.

To write down the matrix $\hat \Lambda$, we assume that the 
 photon propagates in 
the $z$ direction ${\bf k} = k {\bf e}_z$ 
and the external magnetic field is in the $(x,y)$ plane. 
Then 
\be
\mathbf{B}^c = B_0\left(\cos\left(\omega_0 t\right), \sin\left(\omega_0 t\right), 0\right),\;\;\;\; {\mathbf{A}}^q = (A_x^q,A_y^q,0),
\ee
and the matrix $ \hat \Lambda $ takes the form
\begin{equation}
\hat \Lambda = 
\begin{pmatrix}
{\lambda}_{\|}\cos^2(\omega_0 t) + {\lambda}_{\bot}\sin^2(\omega_0 t)   &   \left(\lambda_{\|} - \lambda_{\bot}\right)\cos(\omega_0 t) \sin(\omega_0 t)\\
 \left(\lambda_{\|} - \lambda_{\bot}\right) \cos(\omega_0 t) \sin(\omega_0 t)  &  {\lambda}_{\|}\sin^2(\omega_0t) + {\lambda}_{\bot}\cos^2(\omega_0t)
\end{pmatrix}.\
\label{eq1_1}
\end{equation}
In Eq.(\ref{eq1_1}) we introduced
\begin{equation}
\lambda_{\|}=7\xi k^2 {B_0}^2,\;\; \lambda_{\bot}=
4\xi k^2 {B_0}^2.
%\end{displaymath}\\
%\begin{displaymath}
\end{equation}

To solve Eq.(\ref{eq2}) it is convenient to work in the 
reference frame that rotates with the angular frequency 
$\omega_0$; introducing two vectors 
${\bf n_{\|}} = (\cos(\omega_0 t), \sin(\omega_0 t), 0)$ and 
${\bf n_{\bot}} = (-\sin(\omega_0 t), \cos(\omega_0 t), 0)$, we write
\be
{\bf A}^q = A_{\|} {\bf n_{\|}} + A_\bot {\bf n_{\bot}}.
\label{eq3}
\ee
Substituting the Ansatz Eq.(\ref{eq3}) into Eq.(\ref{eq2}) and keeping 
terms of the first order in $w_0/k$ and $\xi B_0^2$,  we obtain 
the system of equations for $A_{\|,\bot}$
\ba
&& {{\partial}_0}^2A_{\|} - 2\omega_0\partial_0 A_{\bot} 
%- \omega_0^2 A_{\|} 
- \partial_z^2 A_{\|}= \lambda_{\|}A_{\|}, \nonumber \\
&&~ \label{eq4}\\
&& {{\partial}_0}^2A_{\bot} + 2\omega_0\partial_0 A_{\|} 
%- \omega_0^2A_{\bot} 
- \partial_z^2 A_{\bot}= \lambda_{\bot}A_{\bot}. \nonumber
\ea
Eqs.(\ref{eq4}) imply that in the presence of the rotating magnetic field, 
the parallel and transverse components of the photon vector potential 
 are coupled to 
each other; the system of equations that describes their dynamics is 
reminiscent of a coupled harmonic oscillator. 

\section{Solutions for the photon field}

To solve Eq.(\ref{eq4}) we 
make an Ansatz for the photon vector potential
\begin{equation}
A_{\|,\bot} = a_{\|,\bot}^{\pm} e^{i\left(kz - \omega_\pm t\right)}.
\end{equation}
Upon substituting it into Eqs.(\ref{eq4}) and neglecting terms 
${\cal O}(\omega_0^2/k^2)$, we find 
\begin{equation}
\omega_{\pm}^2 = k^2  \left[ 1 
%+ \left (\frac{\omega_0}{k}\right)^2 
-\frac{1}{2}\left(\eta_{\|}+\eta_{\bot}\right) \pm
\frac{1}{2}D^{\frac{1}{2}}\right],
\label{eq6_1}
\end{equation}
where $ \eta_{\|,\bot} = \lambda_{\|,\bot}/k^2 $ and 
\begin{equation}
D = \left(\eta_{\|}-\eta_{\bot}\right)^2 + \frac{16 \omega_0^2}{k^2}.
\end{equation}

The eigenvectors corresponding to $\omega_{\pm}$ 
can be written as
\be
a^+ = 
\left (
\begin{array}{c}
a^+_{\|} \\
a^+_{\bot}
\end{array}
\right )
=
\left (
\begin{array}{c}
i \sin(\delta/2)\\
\cos(\delta/2) 
\end{array}
\right ),\;\;\;\;\
a^- =
\left (
\begin{array}{c}
\cos(\delta/2) \\
i\sin(\delta/2)
\end{array}
\right ),
\ee
where 
\be
\sin \delta = \frac{4\omega_0}{k D^{1/2}},~~~~
\cos \delta = \frac{\eta_{\|}- \eta_{\bot}}{ D^{1/2}}.
\label{eq5_1}
\ee

As can be seen from the above expressions, in the presence of the rotating magnetic field, both the photon vector potential 
and the photon refraction indices $n_{\pm} = k/\omega_{\pm}$,  
depend in a non-trivial 
way on the interplay between the 
 small parameters in the problem -- the ratio of the magnetic 
field angular rotation frequency  and the wave vector of the incoming photon
$\omega_0/k$ and $ (\lambda_{\|}-\lambda_{\bot})/k^2 \sim \xi B_0^2$, 
that describes 
the importance of the non-linear QED effects for  photon propagation. The 
well-known  case of the {\em static} external magnetic field \cite{adler1} 
corresponds 
to $\omega_0/k \ll \xi B_0^2$. However, 
the PVLAS setup corresponds to the {\em opposite} extreme case; 
using values of the external magnetic field, the wavelength of the laser light 
and  the magnetic field angular rotation frequency
reported in \cite{pvlas}, 
it is easy to see that 
$\omega_0/k$ exceeds  
$\xi B_0^2$ by seven orders of magnitude. Hence, 
it is important to analyze the impact of the rotating magnetic field 
on the interpretation of the PVLAS result since it is not obvious that 
the equation for  vacuum birefringence
derived  
in the static approximation for the external magnetic field \cite{adler1}
is  applicable.

To do so,  we write the photon vector potential as a linear combination 
of the normal modes derived earlier
\be
\left ( 
\begin{array}{c}
A_{\|} \\
A_{\bot}
\end{array}
\right ) 
= \sum_{n = \pm}C_n\; a^n\; e^{ikz - i \omega_n t},
\label{eq5}
\ee
where $C_{\pm}$ are the two independent constants to be determined from 
the initial condition.  As the initial condition, we assume that at 
$t=0$ the photon is linearly polarized and its polarization vector 
is at angle $\theta$ relative to  the $x$ axis; this implies  
${\bf A}^q(t=0) = A_0 e^{ikz} \left ( \cos \theta, \sin \theta, 0 \right )$. We then 
find
\ba
&& C_+ = A_0 
\left[ \cos \left ( 
\frac{\delta}{2}   \right ) \sin \theta 
- i \sin \left ( \frac{\delta}{2} \right ) 
\cos \theta  \right ], \nonumber \\
&&~ \label{eq6}\\
&& C_- = A_0 
\left[ \cos \left ( 
\frac{\delta}{2}   \right ) \cos \theta 
- i \sin \left ( \frac{\delta}{2} \right ) 
\sin \theta  \right ]. \nonumber
\ea

We can now combine Eqs.(\ref{eq3},\ref{eq5},\ref{eq6}) to obtain  
the photon vector potential as a function of time. We write
\be
 A_{\|} = A_0 e^{ikz - i\Omega t} \left  ( A_{\|}^{R} + i A_{\|}^{I} \right ),\;\;\;\
A_{\bot} = A_0 e^{ikz - i\Omega t} \left  ( A_{\bot}^{R} + i A_{\bot}^{I} \right ),\;\;\;\ 
\label{eq19}
\ee
where $\Omega = (\omega_+ + \omega_-)/2$ and 
\ba
&&A_{\|}^{R} = \cos \theta \cos \left ( \frac{\Delta \omega t}{2} \right )
+ \sin \theta \sin \delta \sin \left ( \frac{\Delta \omega t}{2} \right ),\;\;
A_{\|}^{I} = \cos \theta \cos \delta \sin \left ( \frac{\Delta \omega t}{2} \right ),
\nonumber \\
&&~ \label{eq20}\\
&& A_{\bot}^{R} = \sin \theta \cos \left ( \frac{\Delta \omega t}{2} \right )
- \cos\theta \sin \delta \sin \left ( \frac{\Delta \omega t}{2} \right ),\;\;\;
A_{\bot}^{I} = -\sin \theta \cos \delta \sin \left ( \frac{\Delta \omega t}{2} \right ), \nonumber
\ea
with $\Delta\omega = \omega_{+} -\omega_{-}$. 
We obtain the photon vector potential by combining Eqs.(\ref{eq19},\ref{eq20}) with 
Eq.(\ref{eq3}). 

We can use those results to discuss important issues for the PVLAS 
experiment. First, we study the polarization of 
the photon as it travels through the external rotating magnetic field.
Recall that at $t=0$, the photon is linearly 
polarized. The density matrix is a convenient tool  to describe the 
photon polarization. 
In particular, we are interested in the Stokes parameter 
$\xi_2$ \cite{ll} that is directly related to the induced ellipticity.
We write $\xi_2$ through the $x$ and $y$ components
of the vector potential and use Eq.(\ref{eq3}) to obtain
\be
\xi_2 = 2 {\rm Im} \left [ \frac{A_y A_x^{*}}{A_0^2} \right ] = 
2 \left [ A_{\|}^R A_{\bot}^I - A_{\bot}^R A_{\|}^I \right ]. 
\ee
Finally, with the help of Eq.(\ref{eq20}), we find
\be
\xi_2 = 2  \cos \delta \sin\left(\frac{\Delta\omega}{2}t\right)
\left[\sin\left(\frac{\Delta\omega}{2}t\right)\cos 2\theta \sin \delta 
- \cos\left( \frac{\Delta\omega}{2}t\right)\sin 2\theta\right].
\label{eq7}
\ee

For the PVLAS setup, the photon spends about $10^{-2}~{\rm s}$ in 
the region with the magnetic field. The magnetic field rotation frequency 
is about $1~{\rm Hz}$. Therefore, $\Delta \omega\; t \ll 1$
 independent of the relationship between 
$\omega_0/k$ and  $\xi B_0^2$, the two 
small parameters in the problem, we can simplify Eq.(\ref{eq7}) by 
expanding in $\Delta\omega\; t$. Through  first order we obtain
\be
\xi_2 \approx  -2  \cos \delta \; \sin 2 \theta\;\; \frac{(\omega_+-\omega_-)t}{2}\;. 
\label{eq8}
\ee
This equation can be further simplified if we use Eq.(\ref{eq6_1})
to derive $\displaystyle \omega_+ - \omega_- = \frac{kD^{1/2}}{2}$, 
trade the time $t$ for the optical 
path $L$ and employ  Eq.(\ref{eq5_1}) to remove $\cos \delta$.
We find
\be
\xi_2 \approx -2 
 \frac{\lambda_{\|} - \lambda_{\bot}}{4k} \sin 2 \theta 
= -2 \; \frac{3\xi k B_0^2}{4} L \sin 2 \theta.
\label{eq9}
\ee

To compute the ellipticity $\epsilon$, we use
\be
\epsilon = \tan(\chi),~~~\xi_2 =  \sin(2\chi).
\ee
Since the ellipticity angle $\chi$ is small, we obtain 
\be
\epsilon \approx \chi \approx \frac{\xi_2}{2} = 
- \frac{3\xi k B_0^2}{4} L \sin 2 \theta.
\label{eq10}
\ee

From Eq.(\ref{eq10}), we observe that 
the induced ellipticity of the photon propagating through the rotating 
magnetic field is independent 
of the magnetic field rotation frequency, in contrast to the fact that 
the solution for the vector potential ${\bf A}$   
exhibits a strong dependence 
on the relationship between 
$\omega_0/k$ and  $\xi B_0^2$.  Hence, the 
analysis of the PVLAS result based on the   equation for 
induced ellipticity derived for static magnetic field \cite{adler1}  
does not introduce 
an error in spite of the fact that the PVLAS operates in the regime which 
corresponds to ``fast'' magnetic field rotation.

Another interesting  question relevant for the PVLAS experiment 
is harmonics generation.
From Eqs.(\ref{eq5},\ref{eq3}), it is clear that the time dependence of 
the photon field is governed by the following factors
\be
{\bf A} \sim e^{-i\omega_{\pm}t} e^{\pm i \omega_0 t},
\label{eq7_1}
\ee
where the last exponential factor comes from the time dependence of 
the vectors ${\bf n_{\|,\bot}}$, introduced in Eq.(\ref{eq2}).
The PVLAS setup corresponds to 
$\omega_0/k \gg \xi B_0^2$. In this limit, the 
photon angular frequencies are 
$\omega_\pm \approx k \pm  \omega_0$. Hence, for the 
PVLAS set up, the propagating 
photon should have components with angular frequencies $k,k \pm 2\omega_0$ 
and no others.  The PVLAS experiment observes $k, k \pm \omega_0, k \pm 2 \omega_0$ but the origin of $k \pm \omega_0$ sidebands is not well understood 
\cite{pvlas,lam}.

\section{Conclusions}

In this paper we have shown that the 
impact of the magnetic field rotation on the 
photon propagation in vacuum depends in a non-trivial way on the 
interplay between  two small parameters, $\omega_0/k$ and  $\xi B_0^2$.
While the well-studied case of a time-independent magnetic field \cite{adler1} 
corresponds to $\omega_0/k \ll \xi B_0^2$, the PVLAS experiment \cite{pvlas} 
operates 
in the regime $\omega/k_0 \gg \xi B_0^2$. 
Nevertheless, it turns 
out that this difference has no bearing on the induced ellipticity 
of the photon; this can be seen from Eq.(\ref{eq10}) where all the 
dependence on the magnetic field angular rotation frequency cancels out.

Our results are in variance with findings reported 
in Ref.\cite{port}. For example, 
 it is claimed in that reference that the rotation of the 
magnetic field generates infinite 
number of higher harmonics in the propagating light beam,  
with frequencies 
$\omega_n = k \pm n \omega_0$, $n =0,1,2,{\rm etc.}$ The difference 
between our computation and that of  Ref.\cite{port}
can be traced back to the incorrect  equation of 
motion for the photon vector potential 
stated in that reference.

{\bf Acknowledgments} 
K.M. is supported in part by the DOE grant DE-FG03-94ER-40833, Outstanding 
Junior Investigator Award and by the Alfred P.~Sloan Foundation. 

{\it Note added} When this paper was being prepared for publication, Ref.\cite{adler2} appeared where the influence of the magnetic field rotation on vacuum 
birefringence was discussed. The conclusion of Ref.\cite{adler2} about the 
independence of the induced ellipticity on the magnetic field rotation 
frequency agrees with ours.

\end{document}